\documentclass[prl,twocolumn,superscriptaddress]{revtex4-2}
\usepackage{amsmath,amssymb,mathrsfs}
\usepackage{graphicx}
\usepackage{colortbl}
\usepackage{braket}
\usepackage{bm}
\usepackage{amsfonts}
\begin{document}
\title{From Complexification to Self-Similarity: New Aspects of Quantum Criticality}

\author{Yang Liu}
\affiliation{Key Laboratory of Polar Materials and Devices (MOE), School of Physics and Electronic Science, East China Normal University, Shanghai 200241, China}

\author{Erhai Zhao}
\altaffiliation{ezhao2@gmu.edu}
\affiliation{Department of Physics and Astronomy, George Mason University, Fairfax, Virginia 22030, USA}

\author{Haiyuan Zou}
\altaffiliation{hyzou@phy.ecnu.edu.cn}
\affiliation{Key Laboratory of Polar Materials and Devices (MOE), School of Physics and Electronic Science, East China Normal University, Shanghai 200241, China}

\begin{abstract}
Quantum phase transitions are a fascinating area of condensed matter physics. The extension through complexification not only broadens the scope of this field but also offers a new framework for understanding criticality and its statistical implications. This mini review provides a concise overview of recent developments in complexification, primarily covering finite temperature and equilibrium quantum phase transitions, as well as their connection with dynamical quantum phase transitions and non-Hermitian physics, with a particular focus on the significance of Fisher zeros. Starting from the newly discovered self-similarity phenomenon associated with complex partition functions, we further discuss research on self-similar systems briefly. Finally, we offer a perspective on these aspects.
\end{abstract}

\maketitle

{\it 1. Introduction}.
Quantum phase transitions (QPTs)~\cite{Sachdevbook} appear at any point of non-analyticity in the ground state energy in the thermodynamic limit. They are the result of the competition between various phases under quantum fluctuations. Understanding quantum phase transitions is of great significance for exploring novel states of matter and exotic excitations.
In addition to the classical mechanism of symmetry breaking, new mechanisms such as deconfined quantum criticality~\cite{Deconfined2004}, long-range entanglement, and symmetry fractionalization~\cite{SPT} provide ways to understand states of matter and phase transitions.
Utilizing nontrivial geometric properties~\cite{diep2013}, beyond-nearest-neighbor coupling effects~\cite{J1J2_1}, and exotic interactions~\cite{Kitaev20062}, model builders aim to introduce frustration to enhance quantum fluctuations. This provides an arena for exploring states of matter, such as spin liquids~\cite{ANDERSON1987,Savary2016,RMP2017SL}, and their associated phase transitions, with the potential to understand the fundamental "holy grail" problem of superconductivity~\cite{RMP06Doping,Chen2020cpl,Wang2024cpl}.
The complexity of most models 
forces researchers to rely on powerful many-body numerical methods such as Monte Carlo sampling~\cite{Sandvik2010} or renormalization group (RG) blocking~\cite{DMRG,TNreview1,TNreview2}. 
However, both theoretically and numerically, the majority of current approaches consider the impacts of thermal fluctuations and quantum fluctuations separately, either starting from finite temperature and then approaching zero temperature, or vice versa. 

In order to depict QPT and their critical behavior more effectively, are there other approaches to better characterize the mutual influence between thermal fluctuations and quantum fluctuations, and utilize numerical experiences to detour the exponential wall?
Meanwhile, in recent years, one has witnessed the rapid development of dynamical quantum phase transitions and non-Hermitian systems incorporating additional degrees of freedom such as real time and dissipation. 
Another important question worth considering is: What intrinsic connections do these scenarios have with the physics of quantum phase transitions in equilibrium systems?

\begin{figure}[t]
  	\centering
  	\includegraphics[width=0.48\textwidth]{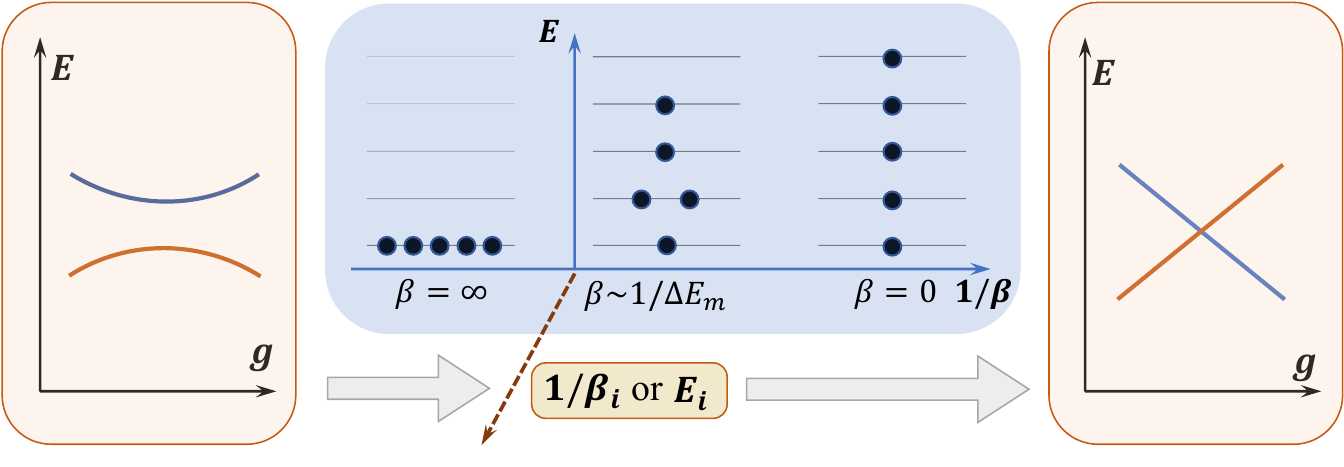}
  	\caption{A schematic graph illustrating the complexification of temperature ($1/\beta$) or energy ($E$) in a finite system. The left box shows that the spectra are analytic and avoid level-crossing between the ground and excited states as the physical parameter $g$ is tuned in a finite system. In the middle box, for a particular $g$, the state distribution, which determines the partition function, changes from fully occupying the ground state to being widely distributed across all states as $1/\beta$ increases from 0 to infinity, passing through a point where the energy level variation is at its maximum ($\Delta E_m$). Due to the complexification of $1/\beta$ or $E$, the analytic partition function collapses at $\beta\sim1/\Delta E_m$, resulting in a level-crossing for a finite system (shown in the right box), which originally only appears at the thermodynamic limit for real $1/\beta$ or $E$.}
  	\label{fig:fig1}
\end{figure}  

The purpose of this short review is to demonstrate that by introducing the complex ingredients of some physical quantities, one can elegantly address both of the aforementioned problems. 
Considering the ratio of the energy difference between quantum states $\Delta E$ to thermal energy $kT$, defined as $X\equiv\Delta E/kT=\beta\Delta E$, with $X\in \mathbb{R}$, as $T$ increases from zero to infinity (or $\beta$ decreases from infinity to zero), the competition between the quantum energy levels and the thermal energy drives the system from the ground state to a more even distribution among different energy states. In between, at a specific inverse temperature $\beta_c$, the energy fluctuation $\sqrt{\overline{\Delta E^2}}$ reaches its maximum $\Delta E_m$, which corresponds to the peak in specific heat. For example, In a system with a second-order phase transition in the thermodynamic limit, this $\beta_c$ represents the critical inverse temperature. However, even in a system without any phase transition, there is still a point where energy fluctuations are maximal. For example, a simple two-level system shows a Schottky peak in its specific heat. At this point, the ratio $X=\beta_c\Delta E_m$ can still be defined, but $\beta_c$ no longer represents a critical quantity.
After analytically continuing the ratio $X$ to the complex plane ($X\in \mathbb{C}$), we analyze different cases. 
First, in the case where $X$ is purely imaginary, it corresponds to a Wick rotation $\beta\rightarrow it$, characterizing the physics of dynamical phase transitions.
Second, when $X$ is a general complex number, one can consider complexification of either $\beta$ or $E$ (Fig.~\ref{fig:fig1}). The former can be used to obtain a complex partition function for a certain static model, while the latter corresponds to energy eigenvalues in non-Hermitian systems. Thus, dynamics and non-Hermiticity are connected to static systems through a simple relationship. 
More importantly, complexification provides a pathway to explore non-analyticity in finite systems. Taking the complex partition function as an example, the original partition function, which describes the number of microscopic states for real $\beta$, collapses at the Fisher zeros. Later, we will illustrate that this singularity can be understood as quantum critical behavior. Quantum fluctuations arising from imaginary part of $\beta$ induce composite effects at certain temperatures, allowing a finite system to exhibit signals of QPT. Therefore, one can characterize the critical behavior of true QPTs by studying the complex partition function and Fisher zeros.

Next, we will provide a detailed overview of significant advancements in complexification in various contexts, including finite temperature phase transitions, dynamical quantum phase transitions, non-Hermitian systems, and quantum phase transitions. Additionally, we will delve deeper into the physical insights gained from a critical self-similarity phenomenon observed through complexification.

{\it 2. Complex Partition Functions and Zeros}.
The concept of complexifying the partition function was first introduced by T. D. Lee and C. N. Yang 70 years ago~\cite{LeeYang1,LeeYang2}. 
They formulated the Lee--Yang Circle Theorem while studying the Ising model in an external magnetic field, which states that for a broad class of systems, including the ferromagnetic Ising model, the partition function's zeros in the complex fugacity plane (Lee--Yang zeros) lie on the unit circle.
In the thermodynamic limit, the clustering of these zeros near the real axis indicates the presence of a phase transition. Specifically, when these zeros pinch the real axis, the system's free energy exhibits non-analyticity, marking the occurrence of a phase transition.
It is noteworthy that this seemingly purely mathematical concept was connected to quantum coherence measurements by Wei and Liu~\cite{Wei2012LYzero}, and was experimentally realized by Peng {\it et al}~\cite{zero_exp1}.
Motivated by Lee and Yang's argument, Fisher extended the concept of complexification to a broader range of partition functions~\cite{fisher1965statistical}. 
\begin{figure}[t]
  	\centering
  	\includegraphics[width=0.4\textwidth]{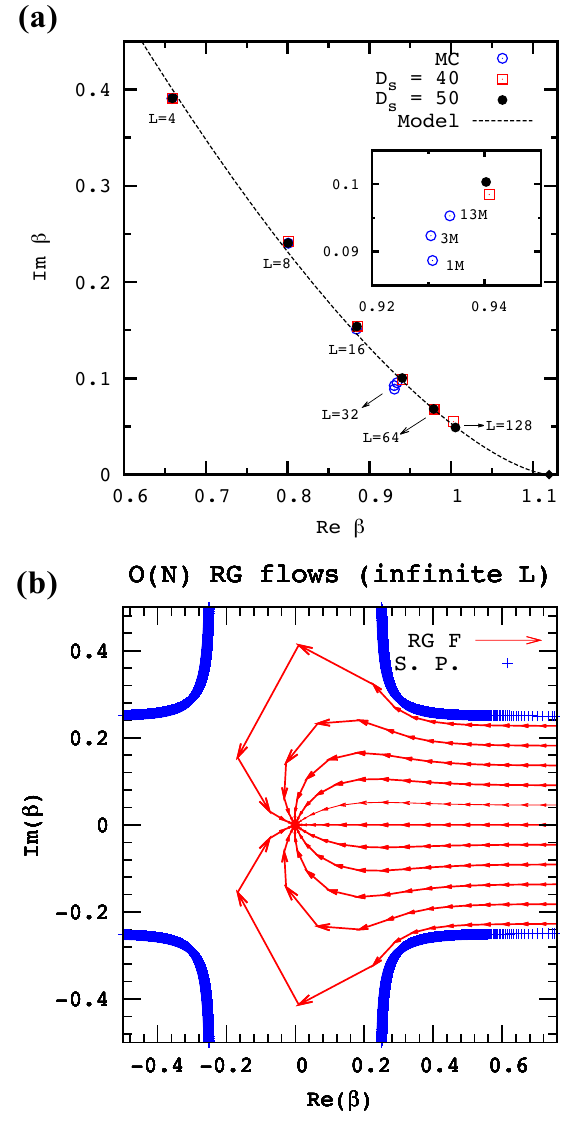}
  	\caption{Fisher zeros characterize phase transitions and RG behaviors.
(a) The scaling behavior of the lowest Fisher zeros to the real axis characterizes the Kosterlitz--Thouless phase transition in the classical XY model as the system size $L$ grows exponentially. Fisher Zero results obtained from HOTRG tensor network methods align with Monte Carlo simulations~\cite{zou2014controlling}. Copyright 2014, American Physical Society.
(b) In the thermodynamic limit of the $O(N)$ model, Fisher zeros (blue) play a crucial role in constraining RG flows (red)~\cite{RGflow2010}. Copyright 2010, American Physical Society.}
  	\label{fig:fig2}
\end{figure}
Instead of the chemical potential or external field, the inverse temperature $\beta$ in the partition function is analytically continued into the complex plane with $\beta=\beta_r+i\beta_i$: 
\[
Z ={\rm Tr}e^{-(\beta_r+i\beta_i)H}.
\]
The resulting zeros of the complex partition function (Fisher zeros) exhibit richer properties. On the one hand, similar to how Lee--Yang zeros characterize phase transitions with an external field, Fisher zeros can be used to study the order of phase transitions, critical exponents, and other thermodynamic properties with the help of finite size scaling~\cite{Biskup2000prl,Biskup2004,Janke2004,Taylor2013}. Specifically, infinite-order topological phase transitions can also be described by the scaling behavior of Fisher zeros [Fig.~\ref{fig:fig2}(a)]~\cite{zou2014controlling}. On the other hand, the distribution of Fisher zeros is highly dependent on the specific model Hamiltonian under consideration and does not exhibit a universal pattern like the Lee--Yang circle theorem. The Fisher zeros of many classical spin models have been studied, revealing various intricate configurations~\cite{Brascamp1974,Wood1985,Kim2008} (For more similar work, refer to the review article~\cite{bena2005statistical}). 
Besides their own scaling behavior, Fisher zeros also play a crucial role in characterizing the intrinsic regulation of RG transformations. The Fisher zeros are situated at the boundary of the attraction basins of infrared fixed points, governing the overarching characteristics of RG flows defined with complex $\beta$ [[Fig.~\ref{fig:fig2}(b)]]~\cite{RGflow2010,Zou2011PRD,Liu2011PRD}.
Due to the intrinsic similarity between lattice field theory and classical spin models, Fisher zeros can also be used to characterize phase transitions in the former~\cite{Barbour1992}.

To locate partition function zeros, initially employing brute force to solve the partition function can provide precise solutions in finite systems. However, achieving thermodynamic limit behavior for complex systems remains challenging through this method. One alternative approach involves using the Ferrenberg--Swendsen reweighting technique to calculate the density of states and then integrating to obtain the partition function~\cite{Daping2012}. 
To mitigate issues stemming from complex values of $\beta$ potentially resulting in negative signs, tensor renormalization group methods~\cite{TRG2007} can be employed. Extending techniques such as higher-order tensor renormalization group (HOTRG)~\cite{XieHOTRG} to the complex plane enables a thorough exploration of partition function zeros~\cite{zou2014controlling}. This methodological extension facilitates the study of critical behaviors beyond real parameters, effectively addressing complexities in thermodynamic systems.

\begin{figure}[t]
  	\centering
  	\includegraphics[width=0.48\textwidth]{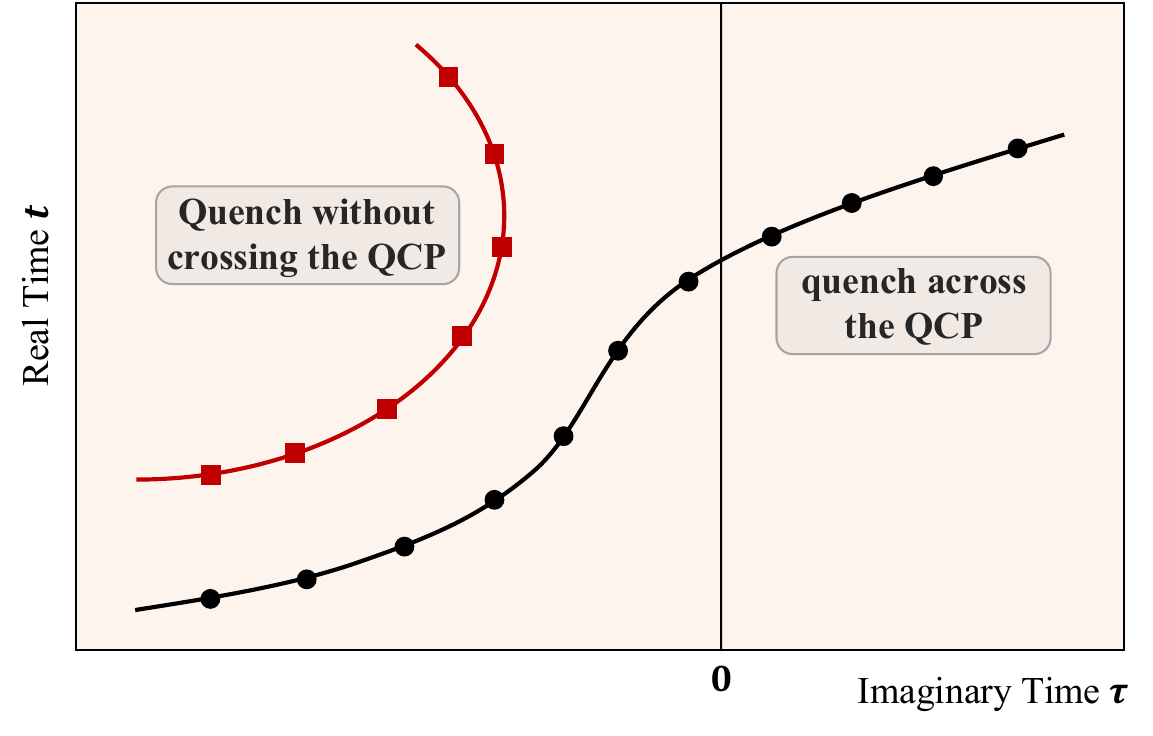}
  	\caption{A schematic graph illustrating a DQPT. When quenching across the QCP of the system with respect to the quenching Hamiltonian, the Fisher zeros (of the boundary partition function) cross the imaginary time line (black). In contrast, when quenching within the same phase, the Fisher zeros remain on one side of the imaginary time line (red). Dots and lines correspond to the Fisher zeros of finite and infinite systems, respectively.}
  	\label{fig:fig3}
\end{figure}  
 
{\it 3. Examples of Complexification on Dynamics and Dissipation.}
Naively, the effect of complexification can be understood through a simple damped oscillation equation: $m\ddot{x} + \gamma\dot{x} + kx = 0
$. The damping and oscillation can be controlled by the complex eigenvalues, which arise from the competition between the damping coefficient $\gamma$ and the elastic parameters $mk$.
Dynamical quantum phase transitions and non-Hermitian systems are two typical quantum analogies of these classical cases.

{\it 3.1 Dynamical Quantum Phase Transitions}.
A general dynamical phase transition that occurs as a result of quenching from a specific initial state holds significant implications in the study of non-equilibrium statistical physics~\cite{Polkovnikov2011,Ye2022SCPMA}. These transitions provide crucial insights into the temporal evolution of complex systems and the underlying mechanisms driving phase changes. Dynamical quantum phase transitions (DQPTs)~\cite{DynamicalQPT2013PRL,Heyl2018} are specific to quantum systems and involve non-analyticities in time-evolved quantities resulting from unitary evolution. A DQPT is typically identified through quantities like the Loschmidt amplitude, and it occurs when the Loschmidt amplitude,
\[
G(t)=\langle\Psi_0|e^{-iHt}|\Psi_0\rangle
\]
exhibits non-analytic behavior at specific critical times. The Loschmidt echo,$L(t)=|G(t)|^2$, provides a measure of how the initial state $|\Psi_0\rangle$ overlaps with its time-evolved counterpart under the Hamiltonian $H$. Non-analyticities in $L(t)$ or the related rate function signal DQPTs. By extending real time $t$ to the entire complex plane, the Loschmidt amplitude can be interpreted as a boundary partition function,
\[
Z(z)=\langle\Psi_0|e^{-zH}|\Psi_0\rangle 
\]
with $z=\tau+it$, and $\tau$ refers to the imaginary time. A key tool in understanding DQPTs is the concept of Fisher zeros, which can be extended to study the boundary partition function in the complex time plane. 
When the system is quenched within the same phase with respect to $H$, the Fisher zeros do not cross the imaginary time axis. However, when the system is quenched across a quantum critical point (QCP), the Fisher zeros cross the imaginary time axis, signaling the occurrence of a DQPT (Fig.~\ref{fig:fig3}). This crossing indicates non-analyticities in the Loschmidt echo at critical times, analogous to the appearance of phase transitions in equilibrium systems. Therefore, complexification provides a powerful diagnostic tool for identifying DQPTs. 
The behavior of Fisher zeros crossing the imaginary time axis is also observed in finite systems, providing a clear indication of the thermodynamic limit results. This feature makes numerical methods such as tensor network techniques particularly useful~\cite{Andraschko2014PRB}.

After the original work detailing the one-dimensional transverse-field Ising model~\cite{DynamicalQPT2013PRL}, this concept has been extensively investigated in other systems~\cite{Budich2016DTOP,DQpotts2017,LiLin2023DQPT}. For example, DQPT has been analyzed in a model with a deconfined quantum critical point~\cite{LiLin2023DQPT}.
Experimental systems for the realization of DQPTs~\cite{ExpDQ2017,Zhang2017nature} allow precise control over quench protocols and measurement of time-evolved quantities. Observing the non-analytic behavior in the Loschmidt echo and identifying critical times consistent with theoretical predictions confirm the occurrence of DQPTs and the relevance of Fisher zeros.

{\it 3.2. Non-Hermitian Systems}.
Non-Hermitian systems~\cite{nonHermitian1}, characterized by the potential for complex energy eigenvalues governed by PT symmetry, have opened up a rapidly advancing field.
They often arise in open quantum systems where there is an exchange of energy or particles with the environment, leading to phenomena such as amplification/dissipation and gain/loss effects, which are caused by the imaginary part of the energy.
Unlike Hermitian counterparts, non-Hermitian systems can manifest unique phenomena such as exceptional points and topology~\cite{ElGanainy2018,Bergholtz2021Exceptional,Liao2021Experimental,Pan2021cp} and the non-Hermitian skin effect (NHSE)~\cite{Lee2016PRL,Yao2018PRL,Li2021cpl,Zhang2022NHSE}.
Besides the NHSE in the Su--Schrieffer--Heeger model~\cite{Yao2018PRL}, this striking phenomenon has been extensively investigated in other contexts, such as those involving hybrid topology~\cite{Lee2019Hybrid,Zhang2021nc,Ou2023prbl,Chen2024cpl} or thermal fields~\cite{Li2019science,Cao2021cp,Huang2023cpl,Mao2024cpl}. 

Although the topological properties of non-Hermitian systems have been widely studied, solving many-body non-Hermitian systems remains extremely challenging. More consideration is needed for the eigenstate or transfer matrix calculation in non-Hermitian systems, such as using biorthonormalization technique~\cite{WangXiang1997,HuZX2024prl,Hu2024DMRGnh},considering gauge transformations~\cite{Tang2023arxiv}, and constructing parent Hamiltonians~\cite{Yang2023Construction}. Experimentally, the implementation of non-Hermitian systems provides a new pathway for exploring complexification, such as Lee--Yang zeros~\cite{Lan2024lynh}.   

\begin{figure}[t]
  	\centering
  	\includegraphics[width=0.4\textwidth]{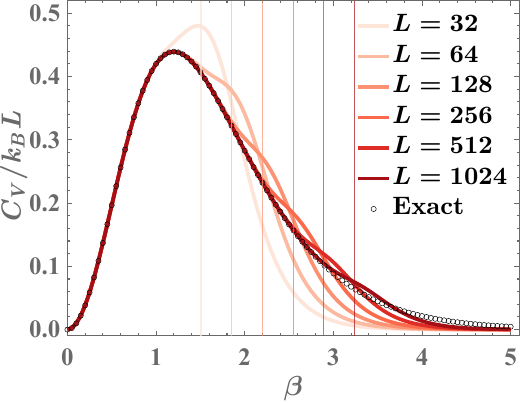}
  	\caption{The specific heat per length $C_V/k_BL$ of the Ising chain exhibits changes as the system size $L$ increases exponentially. The varying shades of red solid lines correspond to $C_V/k_BL$ of finite systems.  The protruding peaks on the specific heat curves corresponds to the $\beta$ precisely matching the maximum real part of Fisher zeros of the system. Black dots without protrusions represent results in the thermodynamic limit, where the zeros form a straight line without boundaries.}
  	\label{fig:fig4}
\end{figure} 

{\it 4. Quantum Criticality from Complexification}.
QPTs occur at zero temperature and appear unrelated to Fisher zeros at finite $\beta$. We first discuss a simple Ising chain without any QPT. Figure~\ref{fig:fig4} shows that 
the specific heat exhibits a peculiar convex peak, coinciding precisely with the real part of the rightmost zero. As the system size $L$ increases, this peak disappears towards large $\beta$.
Correspondingly, the real part of the rightmost zero tends to infinity, and all the Fisher zeros become more closely spaced, forming a continuous line parallel to the real 
$\beta$ axis.
This crossover behavior from finite temperature to zero temperature hints that the Fisher zeros at finite $L$ retain some information at zero temperature in the thermodynamic limit. 
This phenomenon persists even after quantum fluctuations are turned on by applying a transverse field. Therefore, when the transverse field is sufficiently large, the QPT occurring at zero temperature also exhibits signals at finite temperatures.
Indeed, the introduction of imaginary parts of $\beta$ in 
$Z$ causes oscillations between energy levels, revealing quantum fluctuations. Quantitatively, these fluctuations become more pronounced when $\beta_r$ is relatively large.
At the Fisher zeros point, the retained quantum fluctuations cause the microscopic state counting within the framework of statistical mechanics to lose its meaning. Consequently, QPT also manifests effects in this non-analytic region.

\begin{figure}[t]
  	\centering
  	\includegraphics[width=0.48\textwidth]{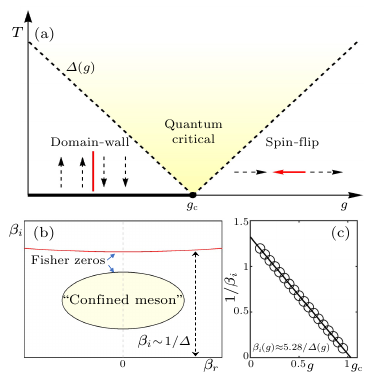}
  	\caption{Fisher zeros of TFIC. (a) Schematic diagram of the quantum critical behavior and the different low-energy excitations on either side of the QCP.
(b) The line of zeros corresponds to excitations of domain walls, while the region enclosed by zero circles represents confined meson excitations.
(c) The imaginary parts of the Fisher zero lines at large $\beta$ provide an energy scale for domain wall excitations~\cite{liu2023CPL}. Copyright 2023, Chinese Physical Society}
  	\label{fig:fig5}
\end{figure}

Exact solutions of Fisher zeros for the transverse field Ising chain (TFIC) confirm that quantum criticality and low energy excitations are indeed related to the Fisher zero patterns (Fig.~\ref{fig:fig5})~\cite{liu2023CPL,Liu2024zero2}. 
In the order phase, the Fisher zero lines correspond to domain wall excitations, and at the QCP, these lines vanish.
The area enclosed by the Fisher zeros hides meson excitations~\cite{liu2023CPL}.
When a longitudinal field is applied, the closed zeros are opened, corresponding to the emergence of $E_8$~\cite{Zou2021PRL,Zhang2020PRB} particles at the QCP.
In the quantum disordered region, the spin flip excitations can be characterized by the scaling behavior of the boundaries of the closed Fisher zero curves.~\cite{Liu2024zero2}. The motion of Fisher zeros in the complex 
$\beta$ plane with respect to the transverse field behaves similarly to the Gr\"uneisen parameter~\cite{Liu2024zero2,Zhang2019prl,Zhang2023cpl}.

\begin{figure}[t]
  	\centering
  	\includegraphics[width=0.48\textwidth]{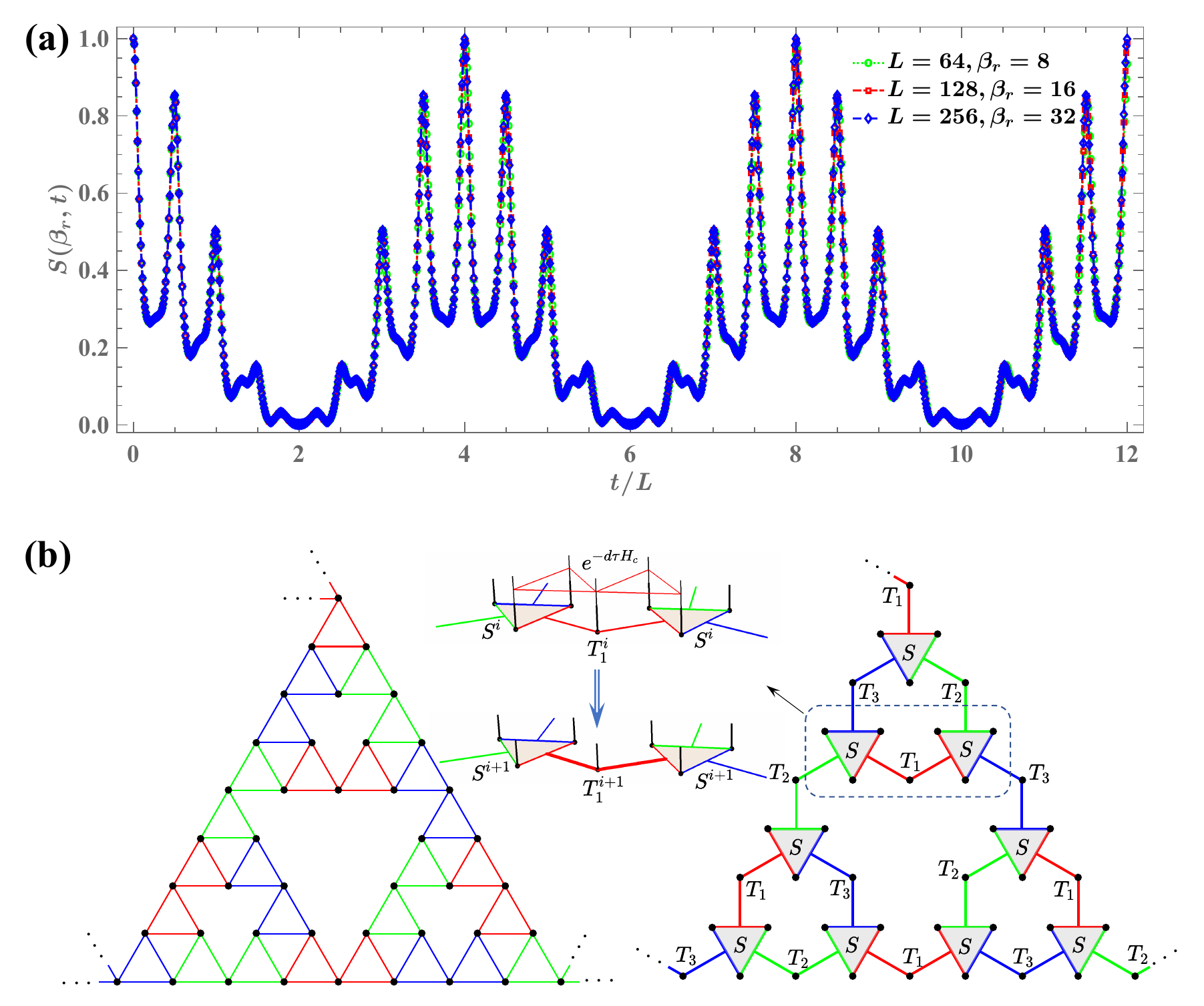}
  	\caption{Two examples of self-similarity. (a) $S(\beta_r,t)$  for different $L$ and $\beta_r$ exhibits self-similarity. (b) The Sierpi\'nski gasket, possessing self-similarity, represents a strongly frustrated structure. The ground state of the spin system on this structure can be determined by using evolution with simplex operators~\cite{Zou2023slcpl}. Copyright 2023, Chinese Physical Society}
  	\label{fig:fig6}
\end{figure}

Another way to determine the quantum criticality from the oscillations of $Z$ is by viewing $Z$ as the survival amplitude of a certain quantum state and calculating the relative norm (or modulus square) of $Z$,
\[
S(\beta_r,t)=\left|\frac{Z(\beta_r,t)}{Z(\beta_r,0)}\right|^2,
\]
where $\beta=\beta_r+it$ is used as it can be treated as a thermofield dynamical problem~\cite{del2017Scrambling}. 
This bears importance in both short-time and long-time dynamics, providing precise information about thermal properties and low-energy excitations, respectively~\cite{Liu2024zero2}. Moreover, at the QPT point, 
$S(\beta_r,t)$ exhibits strong self-similarity [Fig.~\ref{fig:fig6}(a)]~\cite{Liu2024zero2}, providing a new angle for studying quantum criticality.

Firstly, at the QPT point, the divergence of the correlation length leads to self-similar behaviors in various physical quantities, such as wave functions~\cite{Liu2024Lyapunov}. The nearly perfect self-similarity of $S(\beta_r,t)$ at large $\beta_r$ allows one to define scaling variables for RG flows. Starting from zero temperature, the RG flows connect most regions of the complex plane until they are affected by Fisher zeros. 
Secondly, the self-similarity generated by QPT inspires the construction of self-similar structure in real space~\cite{NatPhy2019Smith,PRL2021Jia} to understand potential critical behaviors. From the perspective of frustration, which generates quantum fluctuations, the porous features in a self-similar structure produce larger frustrations than those in triangular or Kagome lattices~\cite{Zou2023slcpl}. The gapless spin liquid results of a spin model on Sierpi\'nski gasket [Fig.~\ref{fig:fig6}(b)] substantiates this viewpoint~\cite{Zou2023slcpl}. 

{\it 5. Summary and Perspective. }
We have reviewed the work related to complexification in phase transitions and dynamics. Complexification offers new research directions and approaches for both finite temperature and quantum phase transitions, and provides a possible pathway to connect dynamical quantum phase transitions and non-Hermitian physics. Notably, the structure of Fisher zeros shows a profound link to quantum critical behavior.
Additionally, self-similarity at the QCP offers insights for discovering new states of matter. 
Despite the important advances in these areas, we have identified several challenges and thought-provoking questions that remain, which are listed as follows:

(I) The experimental observation of Lee--Yang zeros~\cite{zero_exp1,Lan2024lynh} has inspired further efforts to realize Fisher zeros, thereby providing a more direct validation of the fundamental principles of statistical systems.
One proposal involves time-dependent quantum coherence measurement, where the probe-bath interaction Hamiltonian is identical to that of the system~\cite{Wei2014}.
Additionally, $Z$ can be realized using a quantum circuit with both unitary and nonunitary gates~\cite{Liu2024zero2}. The competition between unitary time evolution and quantum measurement in an open quantum many-body system gives rise to an equivalent method for measuring $Z$.
While these approaches are theoretically feasible, their experimental realization still faces numerous technical challenges.

(I$\!$I) Although the correspondence between the critical behavior of the TFIC and Fisher zeros is clear, the physical implications of more interesting models in the complex parameter space remain unknown.
However, the numerical methods for calculating partition functions and their zeros are widely applicable. It is expected that one can explore a wider range of interesting quantum critical behaviors involving various competing states~\cite{Zou2019PRL,Zou2020PRB,Zhu2013PRL,Liu2022Gapless,Zou2023Nearly}.

(I$\!$I$\!$I) While we have noted that both non-Hermitian physics and equilibrium quantum phase transitions employ complexification, their relationship remains somewhat unclear. However, by considering the relationship from the perspective of the partition function, one can choose either $\beta$ or $E$ as complex degrees of freedom. This suggests that a non-Hermitian system $H_{\rm non}$ with complex energy spectrum but real $\beta$ may correspond to a Hermitian system $H$ with complex coupling $\beta_c$: 
\[
Z={\rm Tr}e^{-\beta_r H_{\rm non}}={\rm Tr}e^{-\beta_c H}
\]
As a result, insights gained from studying complex coupling in Hermitian systems, particularly the Fisher zeros, can potentially be applied to non-Hermitian systems. For instance, energy levels in non-Hermitian systems correspond to trajectories in the complex coupling plane. The following important question then arises: Do Fisher zeros have a deeper intrinsic connection to novel behaviors in non-Hermitian systems?

(I$\!$V) It is known that systems containing QPTs exhibit self-similarity of physical quantities at the QCP, suggesting that real self-similar structures may exhibit quantum critical behavior. Controlling the self-similar structure may lead to different outcomes compared to normal structures. For instance, enhanced superconductivity is proposed on fractals~\cite{SC2023fractal}. Therefore, models or materials with self-similar structures have tremendous potential for exploring new states of matter.

{\it Acknowledgements}. We dedicate this work to Professor Tsung-Dao Lee, who passed away on August 4, 2024. We honor his significant contributions to physics, including the pioneering work on the complex partition function with Professor Chen-Ning Yang, and his immense impact on the scientific community in China. His monumental scientific achievements and remarkable character will continue to inspire and guide future generations. We thank Youjin Deng, Shijie Hu, Haijun Liao, Ian McCulloch, Yannick Meurice, Jiansong Pan, Wei Wang, Zhiyuan Xie, and Tao Xiang for helpful discussions. This work was supported by the National Natural Science Foundation of China (Grant No.~12274126). EZ acknowledges the support from NSF grant PHY-206419 and AFOSR
grant FA9550-23-1-0598. 

%

\end{document}